\documentclass[twocolumn,aps,superscriptaddress]{revtex4}
\usepackage{graphicx,amssymb,bm,natbib,amsfonts,amsmath}

\usepackage{graphicx}
\usepackage{appendix}
\usepackage[usenames]{xcolor}
\usepackage[T1]{fontenc}

\begin{document}

\title{Onset of Irreversibility and Chaos in Amorphous Solids Under Periodic Shear}
\author{Ido Regev, Turab Lookman and Charles Reichhardt } 
\affiliation{
Center for Nonlinear Studies and Theoretical Division,
Los Alamos National Laboratory, Los Alamos, New Mexico 87545
} 
\begin{abstract}
An important aspect of the physics of amorphous solids is the onset of irreversible behavior usually associated with yield. Here we study amorphous solids under periodic shear using quasi-static molecular dynamics simulations and observe a transition from reversible to irreversible deformation at a critical strain amplitude. We find that for small strain amplitudes the system exhibits a noisy but repetitive limit-cycle, similar to return point memory~\cite{sethna1993hysteresis}.  However, for large strain amplitudes the behavior becomes chaotic (shows sensitivity to initial conditions) and thus irreversible. We show that the chaotic behavior is a result of the shear band instabilities that arise for large strains and the convective displacement fields they create.
\end{abstract}
\maketitle

\begin{figure}[h]
\includegraphics[width=3.25in]{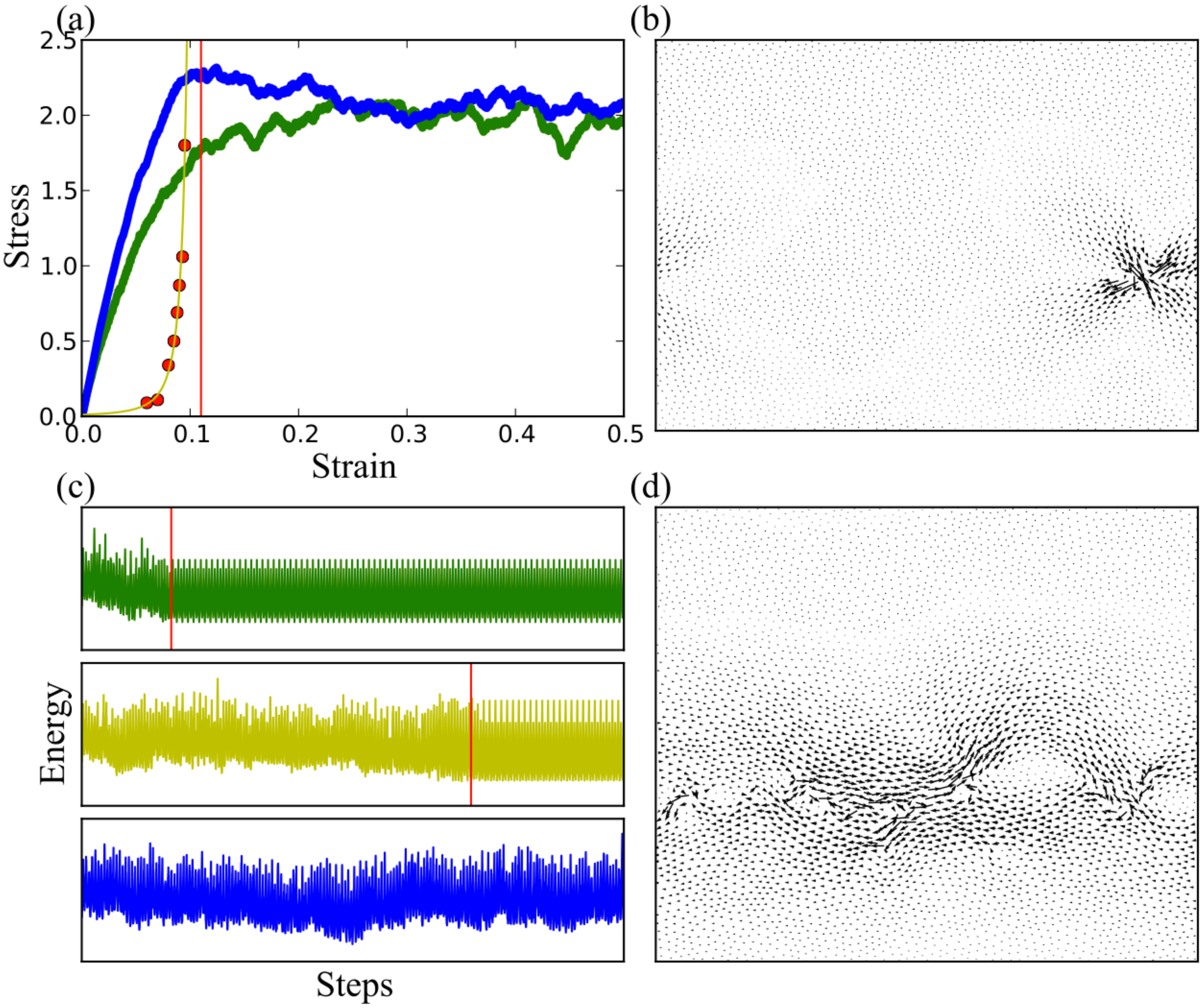}
\caption{
{\bf Plasticity mechanisms and irreversibility in amorphous solids}:
(a) Stress-strain curves from two molecular dynamics simulations for $16384$ particles at the same conditions but with different initial configurations.
Note that while for the blue curve the stress has a maximum which might be used as a definition of yield, for the green curve the onset of plasticity is gradual and there is no distinct point that indicates the onset of plastic deformation.
Red dots represent the typical time scale for relaxation to reversible behavior under oscillator shear. The red line represents the critical strain amplitude for onset of chaos under oscillatory shear (see Fig~\ref{fig:2}) which coincide with yield and the onset of large shear-band like events.
(b) Displacement field that occurs due to localized rearrangements (T1 processes).
(c) Transient behavior of the potential energy before reaching a limit-cycle for three different strain-amplitudes. Red lines are the point at which periodic behavior begins (last one does not reach a limit-cycle at the observed time).
(d) Strain localization (Avalanche) event at an advanced stage of plastic deformation.
}
\label{fig:1}
\end{figure}

Amorphous solids such as plastics, window glass and amorphous metals are an important and ubiquitous form of matter. Industrial processing of such materials commonly involves plastic deformation (for example plastic forming). While a microscopic mechanism of plastic deformation in these materials was identified \cite{Argon,maloney2006amorphous,schall2007structural} the collective behavior on the mesoscale is still being debated. Current theoretical understanding of amorphous solids includes mainly mean-field statistical mechanics theories that are based on the assumption that the behavior is stochastic. Therefore, the dynamics is described in terms of probability distributions which follow the evolution of localized particle rearrangements exhibiting a transition from jammed to flowing behavior~\cite{STZ,SGR1,SGR2,bocquet2009kinetic}. Recent experiments and simulations on superconductor vortices, dilute colloidal dispersions and loosely packed granular materials show that these materials undergo a transition from reversible to irreversible diffusive behavior by varying the strength of an oscillatory external field ~\cite{Sastry,priezjev2013heterogeneous,mangan2008reversible,corte2008random,pine2005chaos,Granular1,Granular2,lundberg2008reversible,schreck2013particle}. In this work we study highly condensed amorphous solids (well above the jamming transition) under oscillatory shear and show that for small strain amplitudes these systems evolve into periodic limit cycles where particles change their positions during the cycle but keep following the same trajectories for consecutive cycles. These reversible rearrangements result in energy fluctuations, which for small strains are completely repetitive which resembles return point memory \cite{sethna1993hysteresis}. Above a critical strain amplitude, the system does not settle into a limit cycle and the motion is chaotic with a positive maximal Lyapunov exponent. This allows us to define a yield point, which can be difficult to determine from a stress strain curve (green curve Fig~\ref{fig:1}). We explain the {\it onset} of chaotic behavior as a result of a convective displacement fields which result from strain localization which generally appears during yield in amorphous solids \cite{chaudhuri2012inhomogeneous,chikkadi2011long,maloney2006amorphous}.  Identifying and understanding the underlying mechanism of chaotic behavior opens the possibility of a quantitative understanding of structural changes that occur in these systems and their relation to the dynamics.

We perform molecular dynamics simulations of a system of N point particles in two or three dimensions interacting via a pair-wise potential (see supplementary material for potential details and simulations in three dimensions) where half the particles are $1.4$ larger than the other half. The sample is kept at a constant number density $\rho=0.75$ which is significantly higher than the jamming transition. Amorphous solids are created by equilibrating the system at a high temperature and than quenching them to zero temperature using a minimization algorithm \cite{FIRE}. Next, the material is  subject to small steps of shear ($\Delta\gamma=10^{-4}$) using the Lees-Edwards boundary conditions. The dynamics under shear is either quasi-static (after each shearing step the energy minimized using a minimization algorithm \cite{FIRE}) or over-damped Brownian motion with zero temperature. The strain is applied in a periodic manner: when a maximal pre-decided strain $\epsilon_{max}$ is reached, the strain is reversed by applying strain steps in the opposite direction. This proceeds until it reaches the negative value of the maximal strain $-\epsilon_{max}$. At this point the strain steps are reversed until the system returns to zero strain, completing the cycle. The cycle is then repeated.  For low strain amplitudes we observe that after a number of oscillations the response of the material becomes completely repetitive (figure~\ref{fig:1}(c)).  However,  the response is not immediately reversible and there is transient non-periodic behavior before the system reaches a stable limit-cycle. The transient times increase with the strain amplitude until the system cannot reach a limit cycle for large strains (see Fig~\ref{fig:1}(c)), similar to what was observed in the shearing of colloidal dispersions \cite{corte2008random}.

\begin{figure}[h]
\includegraphics[width=3.5in]{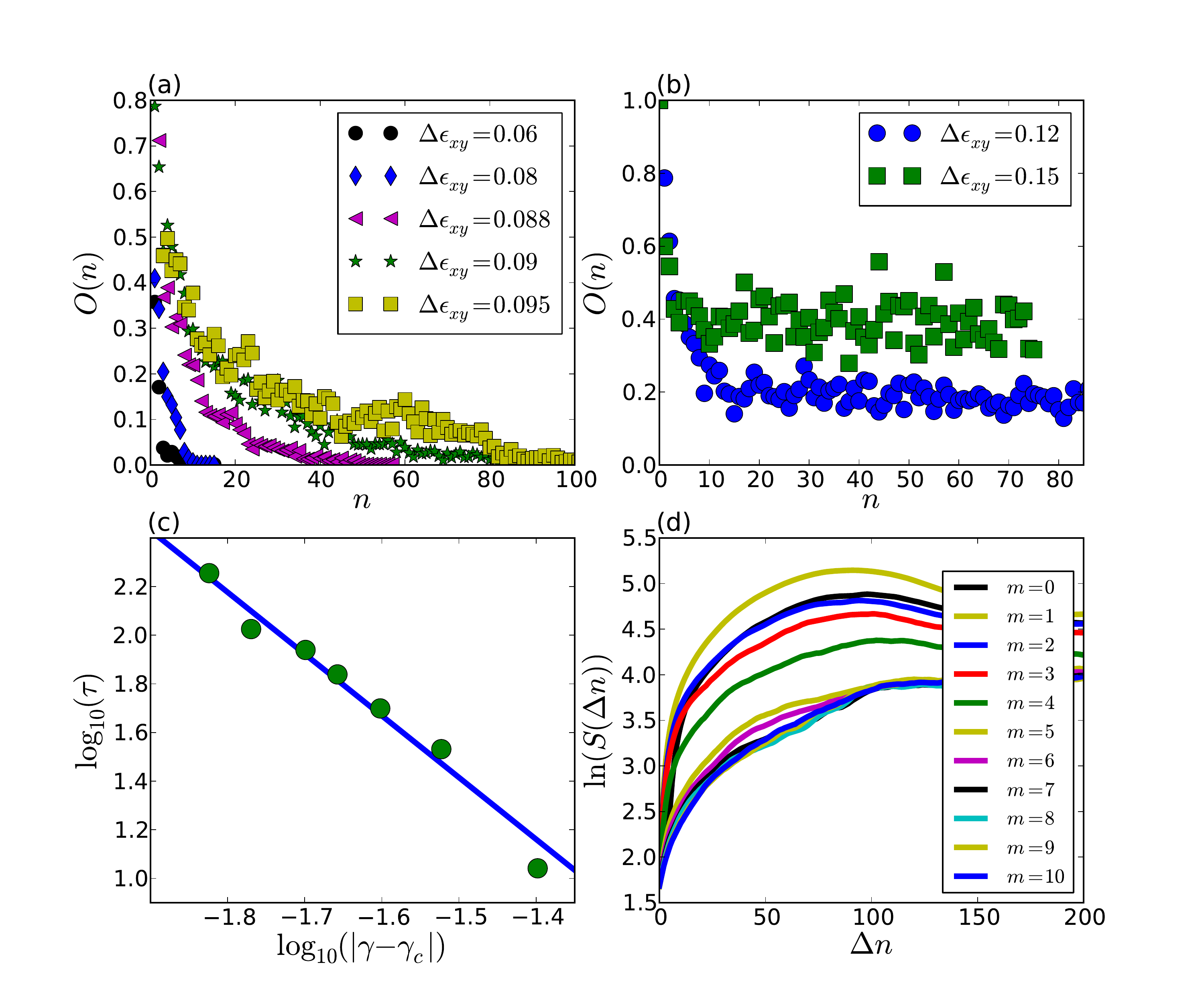}
\caption{ 
{\bf Onset of chaos}:
(a) Overlap function for system size $N=16384$ particles for strains smaller than $\gamma_c=0.11$.
(b) Overlap function for system size $N=16384$ particles for strains larger than $\gamma_c=0.11$
(c) Power-law scaling of the typical time-scale $\tau$ with respect to a critical strain at $\gamma_c=0.11$. The exponent is $\nu\approx -2.54$ and the residuals are $R\approx 2.56\cdot 10^{-4}$.
(d) The function $S(\Delta n)$ as a function of the embedding dimension $m$ for a system of size $N=4096$. For $m>7$ we observe a positive-linear slope which is a signature of sensitivity to initial conditions.
}
\label{fig:2}
\end{figure}
In order to measure the time it takes for the system to reach a periodic limit cycle, we define an overlap (or difference) function for the potential energy $U(t)$:
\begin{equation}
O(n) = \int dt|U(t,n) - U(t,n-p)|\,
\end{equation}
where $n$ is a cycle index and $p$ is the number of cycles before the dynamics repeats itself.
In figures \ref{fig:2}(a) we show this function averaged over $30$ different samples of size $N=16384$, each starting from a different initial condition. For small strains, after a certain number of cycles the difference between the functions $U(t,n)$ and $U(t,n+p)$ approaches $0$. However, for large strains the system does not show tendency towards repetitive behavior and the function approaches a finite asymptotic value (see Fig~\ref{fig:2}(b)).
In most systems that exhibit transient chaos including the Lorenz equations \cite{yorke1979metastable} and certain coupled chaotic maps \cite{tel2008chaotic} the typical time-scale for the transition from chaotic to periodic behavior shows power-law scaling: $\tau \propto (r_1-r)^{\nu}$, where $r$ is a control parameter, $r_1$ is the critical value for the onset of permanent chaos and $\nu$ is a scaling exponent.  In Fig~\ref{fig:2}(c) we can see that in our system the relaxation times follow a power-law with respect to a critical point at $\gamma_c=0.11$ for a system of size $N= 16384$. This is close to the yield strain as estimated from the stress-strain curve in Fig~\ref{fig:1}(a). 

\begin{figure}[h]
\includegraphics[width=3.4in]{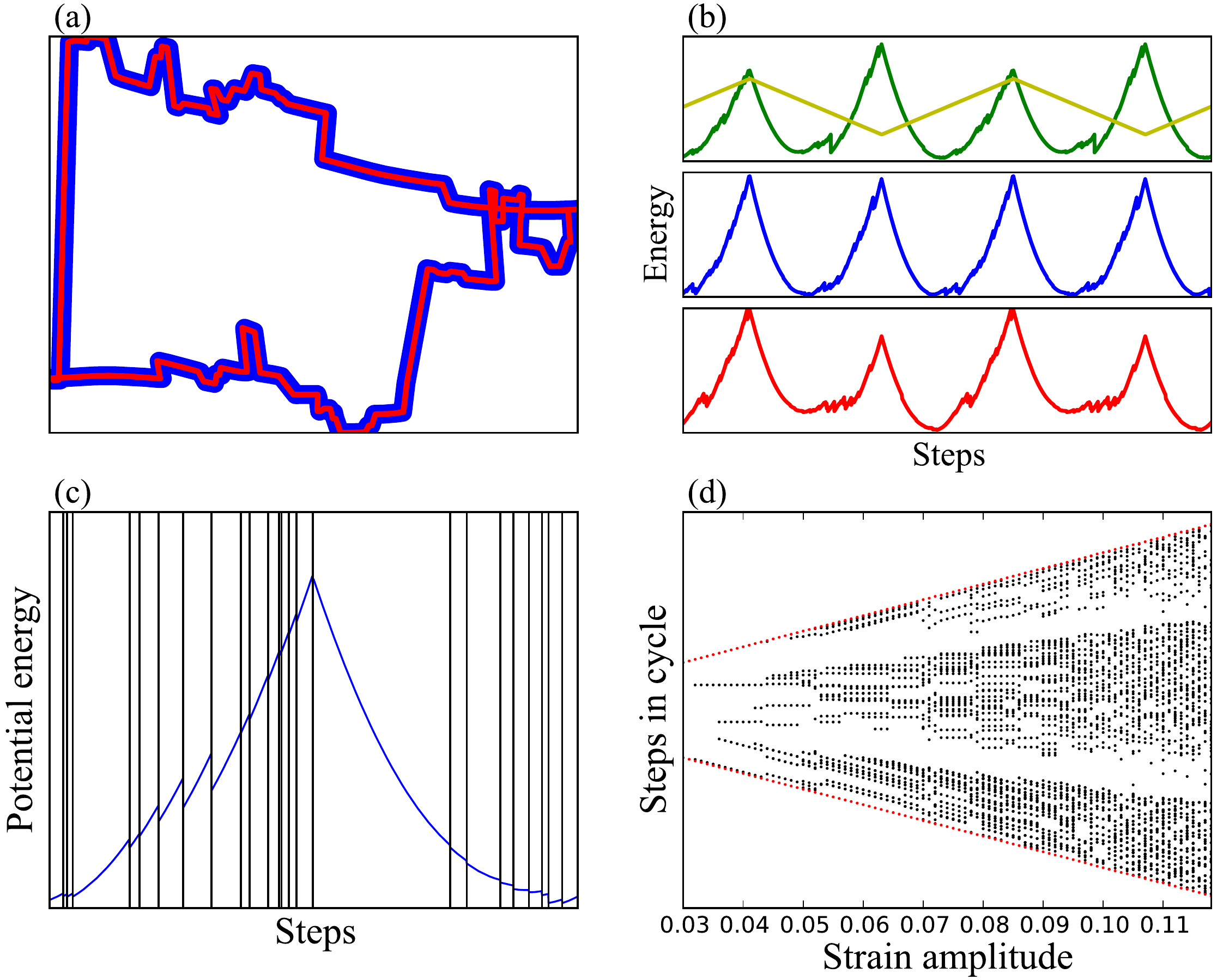}
\caption{
{\bf Behavior of limit cycles}:
(a) Two consecutive trajectories of one particle taken when the system is in a limit-cycle: blue is the first cycle and red is the one just after it. The trajectories
are very similar.
(b) Several different limit-cycles that were obtained using the same control parameters (number of particles, shearing steps, amplitude of shear) but different initial conditions. Even though there are random energy drops, they exactly repeat after one cycle. This behavior resembles the ``return point memory'' effect seen in other random systems such as the Random Field Ising Model \cite{sethna1993hysteresis} . However, for different initial conditions the energy fluctuations are different.
Yellow curve represents the applied strain.
(c) Analysis of one limit cycle with a certain strain amplitude: Energy drops (rearrangement events) are identified and marked as black lines on this curve. The points in the limit cycle where these drops occur are marked as black dots in (d) where time advances from bottom to top. This is repeated for different strain amplitudes (the x-axis in (d)).
(d) A plot of the position of energy drops (rearrangement events) on the limit cycle as a function of the strain amplitude for a system of size $N=1024$. Each vertical column of dots represents the intersection of the black lines from figure (c) with the time axis for a given strain-amplitude. For small strain amplitudes, consecutive limit cycles are similar (energy drops occur at the same times) and change gradually as a function of the strain amplitude. When the strain amplitude approaches a critical value ($\Delta\gamma=0.122$ for this system), consecutive limit cycles become very different from each other indicating the possible existence of riddled basins of attraction where infinitesimally close initial points in state-space can lead to completely different attractors \cite{ott2002chaos}.
}
\label{fig:3}
\end{figure}

In the reversible regimes of dilute colloidal systems the dynamics is quite trivial since particles are no longer in contact \cite{pine2005chaos,corte2008random}. In the highly condensed state studied here, however, particles change positions and rearrange in a non trivial way even during the reversible limit cycles. Typically, this involves a large number of rearrangements of the T1 type (two next nearest neighbors becoming nearest neighbors) which generate elastic-inclusion like displacement fields (see Fig~\ref{fig:1}(b)) and appear as energy drops in the potential energy time-series. This repetitive behavior can also be observed when one the trajectory of any single particle over consecutive cycles (Fig~\ref{fig:3}(a)). In Fig~\ref{fig:3}(b) we see three different limit cycles all simulated with the same system size and strain amplitude but with different initial conditions: we observe that while the period is the same, the details of the cycles (energy fluctuations) depend on the initial configuration. Repetitive fluctuating energy drops are familiar in certain spin-systems where they are known as ``return point memory''~\cite{sethna1993hysteresis}. 

In Fig~\ref{fig:3}(d) we show a plot of the location of the energy drops (black dots) inside a limit cycle at different strain amplitudes starting from a single initial condition. We observe that for small strains, limit-cycles that start from the same initial condition are similar to each other and an increase of the strain amplitude changes the limit-cycles in a gradual manner. However, for large strains small increments in the strain amplitude result in a completely different limit-cycle. We believe that this is a manifestation of the coexistence of many different limit-cycles which occupy different parts of the state-space where  infinitesimally close initial points in state-space can lead to completely different attractors \cite{ott1994blowout}. 
\begin{figure}[h]
\includegraphics[width=3.4in]{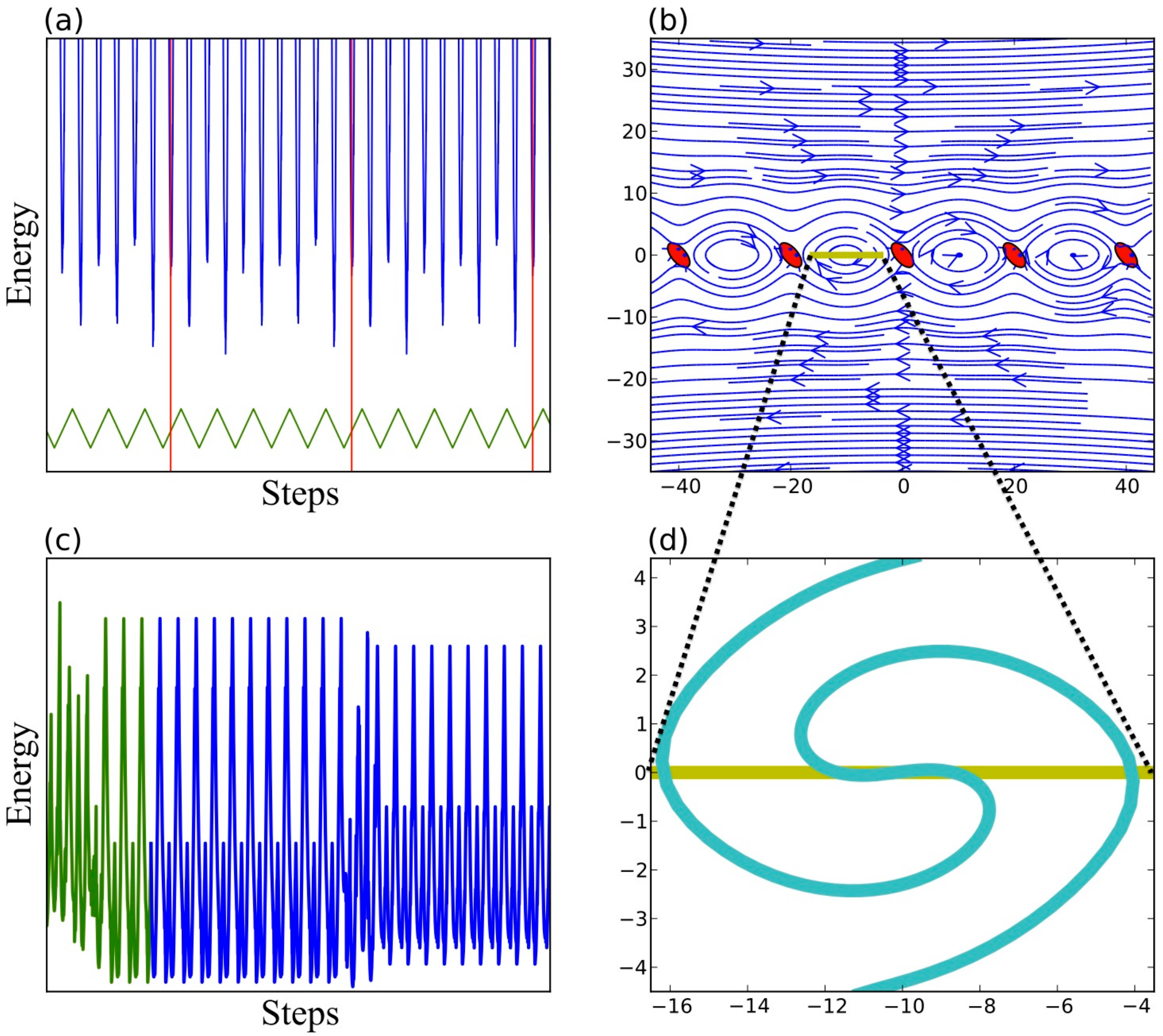}
\caption{
{\bf Routes to chaos}:
(a) Periodic limit cycles with period $5$ at strain amplitude $\Delta\gamma=0.09$. The green curve is the applied strain. Red curves represent the start and the end of a cycle. 
(b) Displacement field due to an array of Eshelby inclusions arranged in the configuration discussed in  \cite{maloney2006amorphous,procaccia1,procaccia2}. Compare the displacement field to Fig~\ref{fig:1}(d).
(c) Effect of thermal noise: System relaxes into a limit-cycle after initial shear (green). It is then subject to the shear accompanied by a small Langevin noise. After some time it ``hops'' to another limit-cycle. This also shows that the limit-cycles can exist with thermal fluctuations.
(d) ``Stretch and fold'' effect: The cyan curve is the result of applying the displacement field from Fig~\ref{fig:4}(b) on a horizontal material line (yellow) $1000$ times. The stretching and folding of material lines causes mixing, sensitivity to initial conditions, chaos and give rise to the diffusive behavior observed in \cite{Sastry}.
}
\label{fig:4}
\end{figure}
While the limit cycles that are shown in Fig~\ref{fig:3}(b) repeat themselves after one cycle, for large strain amplitudes in the larger systems that were studied ($N= 16384 $), we observed cycles that repeat themselves after $2,3,4$ and $5$ cycles (see Fig~\ref{fig:4}(a)). In some dynamical systems, chaos appears due to ``period doubling bifurcations'' so that for certain values of the control parameter the period of the limit-cycle doubles. In these systems a succession of period doubling bifurcations lead to an infinite period and chaos. While we observe periods larger than one, we do not observe a period-doubling cascade (which is a possible route to chaos). It is still possible that such a cascade exists but is obscured by the existence of riddled basins of attraction. In figure~\ref{fig:4}(c) we show the effect of applying noise to a system that is already in a limit-cycle. After a few cycles the system escapes from the initial limit-cycle and settles in a different limit-cycle. 

The most obvious indication of chaotic behavior is the existence of a positive Lyapunov exponent. This is a signature of sensitivity to initial conditions:  trajectories starting from close-by initial conditions diverge exponentially \cite{ott2002chaos}. We check for sensitivity to initial conditions by analyzing the potential-energy time-series using the embedding algorithm of Kantz and Schreiber (see \cite{kantz1997nonlinear} and supplementary material). In this algorithm one calculates the function $S(\Delta n)$ for a guess of the embedding dimension $m$ which is related to the dimension of the space in which dynamics occur (this space is usually referred to as ``the attractor''). This function should exhibit linear behavior for a system with sensitivity to initial conditions. In figure Fig~\ref{fig:2}(d) we show the behavior of the function $S(\Delta n)$ (defined in supplementary material) calculated from the potential energy time-series for a system of size $N=4096$ particles. For embedding dimension  $m>7$, a linear regime appears. Increasing the embedding dimension further does not change the overall behavior of the function. A linear regime with a positive slope is related to the existence of a positive maximal Lyapunov exponent and sensitivity to initial conditions. The positive Lyapunov exponent also implies that the system is ergodic at this stage.

Having established that our amorphous solids simulations show a transition from repetitive to chaotic behavior, we proceed to explain the origin of this chaotic behavior. In figure~\ref{fig:1}(a) we see that after the critical point the stress-strain curve is not smooth anymore but is jugged with large avalanche-like events which result from strain localization \cite{chaudhuri2012inhomogeneous,chikkadi2011long,maloney2006amorphous}. The strain localization is a result of the appearance of a linear array of a sub-extensive number of localized plastic events \cite{procaccia1,procaccia2,chikkadi2011long,maloney2006amorphous}. In Fig~\ref{fig:4}(b) we plot the displacement field that results from such an array of localized plastic events that are represented by an Eshelby inclusion \cite{chaudhuri2012inhomogeneous,chikkadi2011long,maloney2006amorphous,schall2007structural} (compare to the displacement field observed in the simulation in Fig~\ref{fig:1}(d)). The most obvious feature of this field is the convective pattern which appears in between the inclusions. In Fig~\ref{fig:4}(d) we show the effect of applying the displacement field to an initially straight line. After $1000$ iterations the line stretches and folds considerably. This kind of stretching and folding dynamics gives rise to mixing and a finite Lyapunov exponent (sensitivity to initial conditions) \cite{ottino1989kinematics}.
\\

In summary, we have examined the response of a simulated amorphous solid to oscillatory shear and found that for small strain amplitudes the response becomes repetitive after a transient. At large strain amplitudes however, the response becomes chaotic. We establish that there is a ``transition to chaos'' from the periodic to chaotic dynamics which involves a power-law divergence of the time it takes the system to reach a repetitive steady-state. We explain the onset of chaotic dynamics as a result of the displacement fields generated by strain localization. Our results may be verified by experiments on bulk metallic glasses as well as colloidal amorphous solids subject to slow oscillatory shear. Mean field theories, such as the Shear Transformation Zone theory \cite{STZ} show a dynamical transition between jammed and flowing behavior, however, some of the results that we show such as return point memory, cannot be captured by a mean field theory. 

{\bf Acknowledgments}
We would like to thank Paul Chaikin, Colm Connaughton, Bob Ecke and Nicholas Ouellette for useful discussions. Simulations were run on the ``Conejo'' supercomputer at LANL. This work was carried out under the auspices of the U.S. Department of Energy at Los Alamos National Laboratory under Contract No. DE-AC52-06NA25396.

\bibliographystyle{apsrev}
\bibliography{TransitionToChaosLetter} 

\subsection{Supplementary Material}

{\bf Two Dimensional Potential details}:
For the two dimensional system we use the potential:
\begin{equation}\label{potent2}
U(r) =
\left\{
\begin{array}{l}
\epsilon\left[\left(\frac{\sigma}{r}\right)^{12} -
\left(\frac{\sigma}{r}\right)^6 + \frac{1}{4} - h_0 \right]  \ ,\quad  r \le \sigma x_0 \\
\epsilon h_0P\left(\frac{\frac{r}{\sigma} - x_0}{x_c}\right)  \ , \quad \sigma x_0 < r  \le \sigma (x_0 + x_c)  \\
0 \ , \quad  r > \sigma(x_0 + x_c) \ ,
\end{array}
\right.
\end{equation}
which was developed in \cite{lerner2009locality} and consists of the repulsive part
of the standard Lennard-Jones potential, connected via a hump to a region that is smoothed continuously to zero. 
The point $x_0$ is the position at which the LJ potential is minimal, $x_0 \equiv 2^{1/6}$, and the position where the
potential vanishes is $\sigma(x_0+x_c)$. The parameter $h_0$ determines the depth of the minimum. The polynomial $P(x)$ is chosen
as \begin{equation}\label{defineP}
P(x) = \sum_{i=0}^{6}A_i x^i\ .
\end{equation}
with the coefficients given in table \ref{table}.
\\
\begin{table}
\begin{tabular}{|c|c|}
\hline
$A_0$&-1.0\\
$A_1$&0.\\
$A_2$&1.785826183464224\\
$A_3$&28.757894970278530\\
$A_4$&-81.988642011620980\\
$A_5$&76.560294378549440\\
$A_6$&-24.115373520671220\\
\hline
\end{tabular}
\caption{The coefficients in Eq. (\ref{defineP})}
\label{table}
\end{table}

{\bf Two Dimensional Potential Details and Results}
To supplement the simulations in two dimensions we also run simulations of a binary mixture (1:1.4) of repulsive soft spheres using the potential $U(r)\propto\frac{1}{r^{12}}$ in three dimensions. We apply periodic quasi-static shear in the same manner as before. For small strains we see that the dynamics settles into a limit-cycle:
\begin{figure}[h]
\includegraphics[width=3.5in]{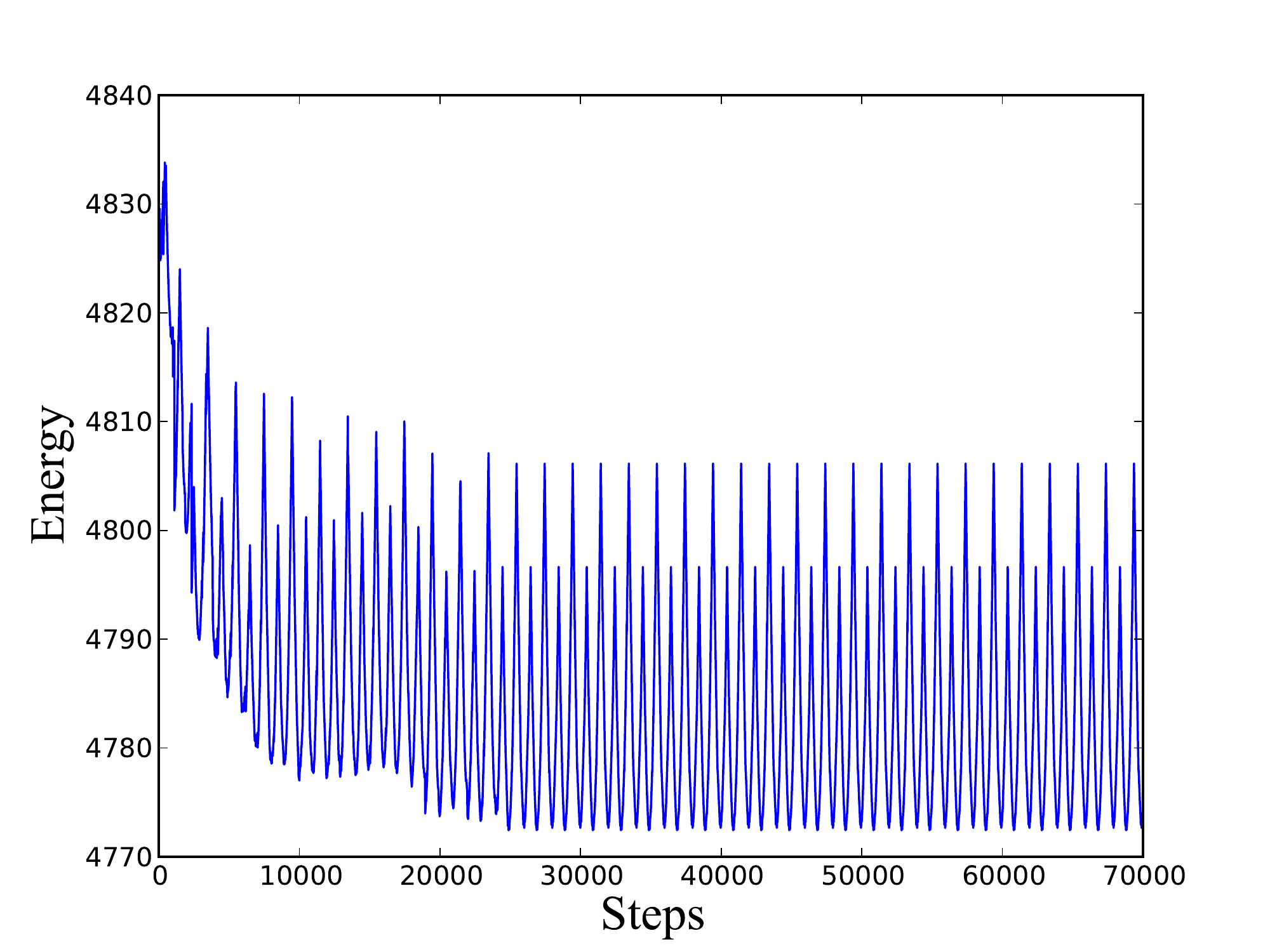}
\caption{Three dimensional soft spheres. Maximal strain is $\Delta\gamma=0.05$.
}
\label{fig:5}
\end{figure}
\begin{figure}[h]
\includegraphics[width=3.5in]{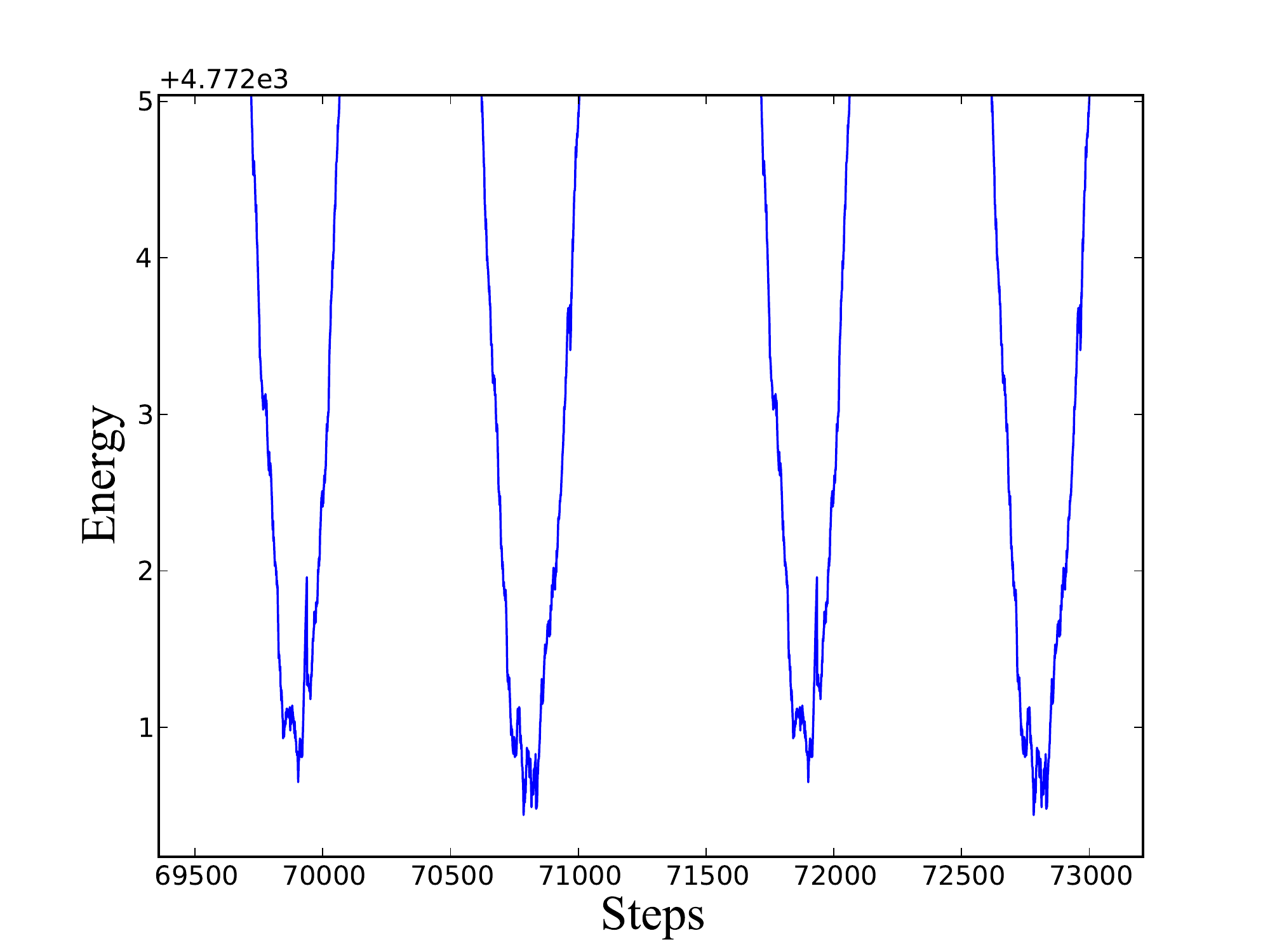}
\caption{Zoom in of previous figure (Fig~\ref{fig:5}).
}
\label{fig:6}
\end{figure}

{\bf Time Series Analysis}:
The method \cite{kantz1997nonlinear} involves the reconstruction of a multi-dimensional state-space from a time-series (in our case, the potential energy):
\begin{equation}
{\bf s}_n = (s_{n-(m-1)\tau},s_{n-(m-2)\tau},...,s_{n-\tau},s_n)
\end{equation}
where $\tau$ is a delay time and $m$ is an embedding dimension (the guessed dimension of the attractor).
For a given $\tau$ and $m$ the algorithm finds points which are close to each-other in embedding space and measures the distance between them as a function of time ($n$). Let $\mathcal{W}$ be a set of delay reconstruction points ${\bf s}_k$ selected at random from the trajectory such that they approximate the true probability distribution. Then $|\mathcal{W}|$ is the number of members in $\mathcal{W}$. The set of points in an $\epsilon$ neighborhood of ${\bf s}_k$ is denoted by $\mathcal{U}_k$. We define:
\begin{equation}
S(\Delta n) = \frac{1}{|\mathcal{W}|}\sum_{k\in \mathcal{W}} \ln{\left(\frac{1}{|\mathcal{U}_k|}\sum_{\ell\in\mathcal{U}_k}|{\bf s}_{k+\Delta n} - {\bf s}_{\ell + \Delta n}|\right)}
\end{equation}
For a chaotic time-series this function shows linear behavior for small $\Delta n$.

\end{document}